\documentclass[aps,prl,twocolumn,showpacs,superscriptaddress,groupedaddress]{revtex4}  
\usepackage{graphicx}  
\usepackage{dcolumn}   
\usepackage{bm}        
\usepackage{amssymb}  
\usepackage{caption}
\usepackage{refstyle}
\usepackage{amsmath}
\usepackage{blindtext}
\usepackage{hyperref}
\begin{document}



\title{Comment on ``Revision of Bubble Bursting: Universal Scaling Laws of Top Jet Drop Size and Speed''}

\author{{\bf J.M. Gordillo}, Departamento de Ingener\'ia Aeroespacial y Mec\'anica de Fluidos, Universidad de Sevilla, Spain\\{\bf J. Rodr\'iguez-Rodr\'iguez}, Grupo de Mec\'anica de Fluidos, Universidad Carlos III de Madrid, 28911, Legan\'es, Spain}



\maketitle

In a recent Letter \cite{PRL2017} Ga\~n\'an-Calvo presents scalings, in very good agreement with experiments, for the velocities and diameters of the first jet drops produced after bubble bursting. It is the purpose of this comment to show that the physical arguments given to explain such scalings are not consistent with theoretical, experimental and numerical evidences reported in the literature, and confirmed by our own simulations. Indeed, the low-Oh limit of Eq. (7) in \cite{PRL2017} expresses that the jet velocity is $V\propto V_c\,Oh^{1/2}$, with $V_c\propto \sqrt{\sigma/(\rho\,R_0)}$ the capillary velocity and $Oh=\mu/\sqrt{\rho\,R_0\,\sigma}$ the Ohnesorge number. This result, combined with Eq. (4) in \cite{PRL2017} yields $V'=V_c$. In addition, the justification of the scalings in \cite{PRL2017} rests on the assumption that all terms in the momentum equation are of the same order of magnitude at the instant of jet ejection, $\rho\,V^2/L\sim \mu V'/L^2\Rightarrow \rho\,V^2\sim \mu V'/L$ from which, since $V\propto V_c\,Oh^{1/2}$ and $V'=V_c$, it is deduced that the unknown length $L$ coincides with the initial radius of the bubble, namely, $L\propto R_0$. In \cite{PRL2017}, the jet velocity $V$ is deduced from the mass balance $V\,R^2\propto V'\,L\,R$, a fact meaning that the proposed physical mechanism assumes that the fluid entering the jet comes from a region of width $L\propto R_0$ of velocity $V'\propto V_c$. Thus, it is hypothesized in \cite{PRL2017} that viscosity sets in motion a region of width comparable to the initial radius of the bubble with a velocity comparable with the capillary velocity $V_c$, a fact which is in contradiction with the well known results by \cite{Moore}, where it is shown that the width of the region where vorticity produced by viscous effects to comply with the shear-free condition at the interface is concentrated in a boundary layer region of width much smaller than $R_0$. Indeed, the numerical results in figure 1 illustrate that the thickness of the boundary layer formed in the high curvature region located at the base of the ejected jet, which depends on the Ohnesorge number, is far smaller than $R_0$.
\begin{figure}
\includegraphics[height=4.5cm]{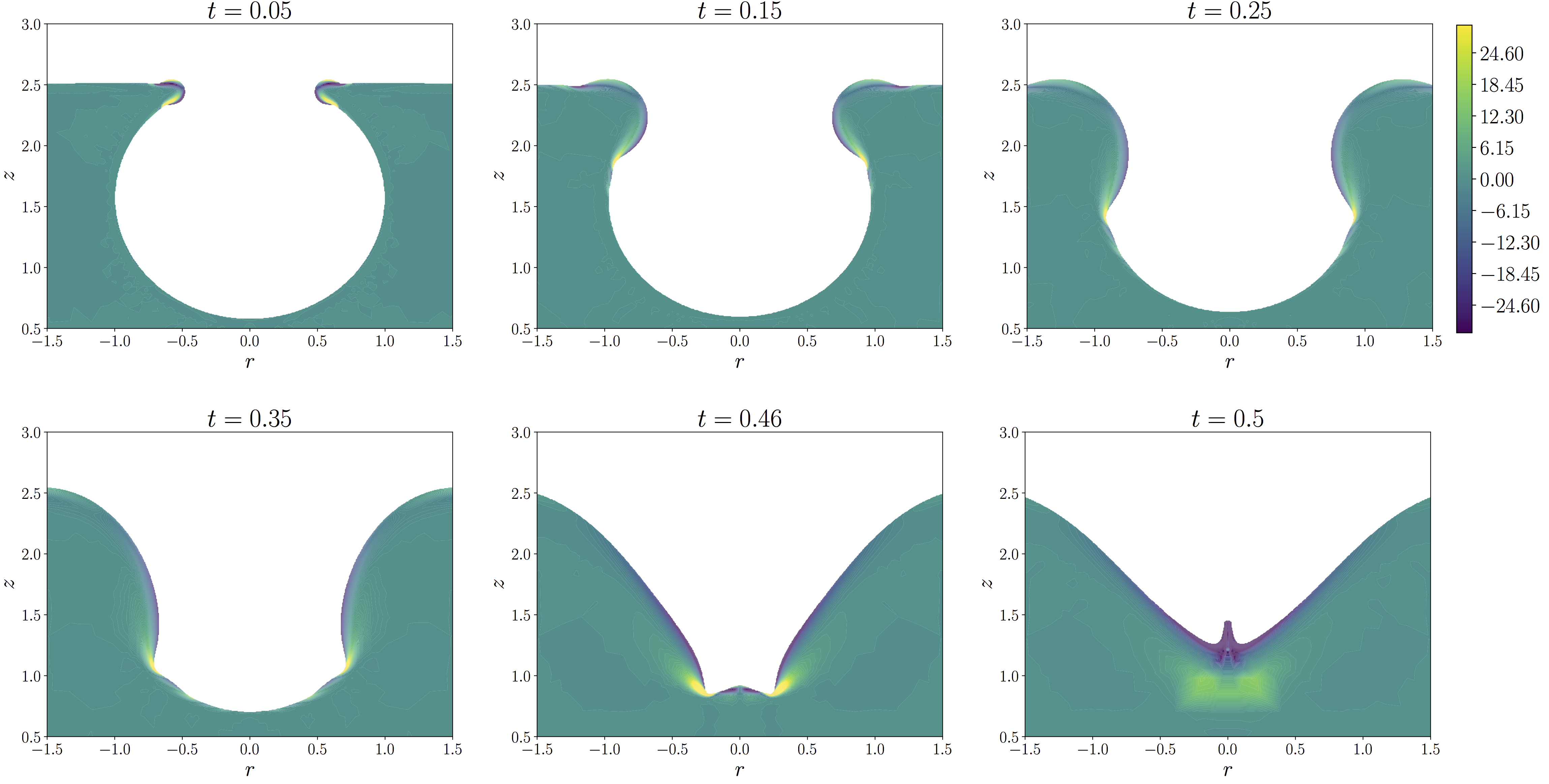}
\caption{\label{fig:vorticity}Vorticity contours for a capillary wave generated upon the recoil of the bubble cap and that focuses at the apex of the cavity leading to the formation of a Worthington-like jet. It is worth remarking that the thickness of the boundary layer induced by the wave propagation is of the order of the wavelength or amplitude, and much smaller than the cavity size. This case corresponds to $Bo = 0.05$ and $Oh = 10^{-2}$.}
\end{figure}\\

Contrarily to the physical mechanism presented in \cite{PRL2017}, which attributes to viscosity the origin of the jet generated after the bursting of a bubble resting on a free interface, in \cite{PRLBursting} we reveal that such jet emerges as a consequence of a purely inertial mechanism, triggered by the presence of the fastest capillary wave which, once generated during the rim retraction process, propagates towards the apex of the air cavity \cite{Duchemin2002}, breaking the self similarity of the inertio-capillary collapse of the void \cite{Zeff,Sierou,Deike,Thoroddsen_2018}. However, viscosity plays a role in the selection of the wavelength breaking the self-similarity, and thus in the modulation of the jet's initial velocity.\\


Funding from MINECO under Projects DPI2017-88201-C3-1-R and DPI2017-88201-C3-3-R is acknowledged. We thank our colleagues and friends for useful suggestions.


\end{document}